\documentclass[
a4paper, 
prb,
twocolumn, 
superscriptaddress, 
amsmath, 
amssymb, 
showpacs
]{revtex4}

\usepackage{graphicx}
\usepackage{dcolumn}
\usepackage{bm}
\usepackage{ulem}
\usepackage{units}

\newcommand{\stras}{
  \affiliation{
  Institut de Physique et Chimie des Mat\'eriaux de Strasbourg, 
  UMR~7504 (ULP-CNRS), BP 43, F-67034 Strasbourg Cedex 2,
  France
  }
}
\newcommand{\augs}{
  \affiliation{
  Institut f\"ur Physik,
  Universit\"at Augsburg, D-86135
  Augsburg, Germany
  }
}

\begin{document}
\title{Surface plasmon in metallic nanoparticles: renormalization effects due to
electron-hole excitations}
\author{Guillaume Weick}
\email{Guillaume.Weick@ipcms.u-strasbg.fr}
\stras \augs
\author{Gert-Ludwig Ingold}
\augs
\author{Rodolfo A.~Jalabert}
\stras
\author{Dietmar Weinmann}
\stras

\date{\today}

\begin{abstract}
The electronic environment causes decoherence and dissipation of the 
collective surface plasmon excitation in metallic nanoparticles. We show that 
the coupling to the electronic environment influences the width and the position of the 
surface plasmon resonance. A redshift with respect to the classical Mie 
frequency appears in addition to the one caused by the spill-out of the electronic 
density outside the nanoparticle. We characterize the spill-out effect
by means of a semiclassical expansion and obtain its dependence on temperature
and the size of the nanoparticle. We demonstrate that both, the spill-out and the 
environment-induced shift are necessary to explain the experimentally observed 
frequencies and confirm our findings by time-dependent local density approximation 
calculations of the resonance frequency. The size and temperature dependence of the
environmental influence results in a qualitative agreement with pump-probe 
spectroscopic measurements of the differential light transmission.
\end{abstract}


\maketitle

%===========================================================================
%===========================================================================
%===========================================================================
%===========================================================================
\section{Introduction}
One of the most prominent features of a metallic nanoparticle subject to an
external driving field is a collective electronic excitation, the so-called 
{\it surface plasmon}. \cite{deheer, brack, kreibig_vollmer} Since 
the first spectroscopic measurement of the related resonance in the absorption 
cross section of free sodium clusters, \cite{deheer87} much progress has been 
made in the characterization of this collective resonance, both experimentally 
\cite{deheer, brechignac92, haberland, brechignac, reiners_cpl, reiners} and 
theoretically. \cite{brack, ekardt, barma, yannouleas, bertsch, madjet} The proposed 
application of metallic nanoparticles \cite{boyer} or nanocrystals \cite{dahan} 
as markers in biological systems such as cells or neurons renders crucial the 
understanding of their optical properties. 

The first experiments have been made on ensembles of nanoparticles, where the 
inhomogeneous broadening of the resonance resulting from the size dependence 
of the resonance frequency masks the homogeneous linewidth. \cite{lamprecht, 
stietz, bosbach} In order to gain detailed information on the collective resonance, 
considerable effort has lately been devoted to the measurement of single-cluster optical 
properties. \cite{klar, sonnichsen, arbouet, dijk, berciaud}
The possibility of overcoming the inhomogeneous broadening resulted in a renewed 
interest in the theory of the optical response of metallic clusters. 

From a fundamental point of view, surface plasmons appear as interesting  resonances 
to study given the various languages that we can use for their description, which
are associated with different physical images. At the classical level, a nanoparticle 
can be considered as a metallic sphere of radius $a$ described by a Drude dielectric function 
$\epsilon(\omega)=1-\omega_{\rm p}^2/\omega(\omega+{\rm i}\gamma_{\rm i})$, 
where $\omega_{\rm p} = \sqrt{4\pi n_{\rm e} e^2/m_{\rm e}}$ is the plasma
frequency, $\gamma_{\rm i}^{-1}$ the relaxation or collision time, while $e$,
$m_{\rm e}$ and $n_{\rm e}$ stand for the electron charge, mass and bulk density, 
respectively. Classical electromagnetic theory for a sphere in vacuum yields a
resonance at the Mie frequency $\omega_{\rm M}=\omega_{\rm p}/\sqrt{3}$. 
\cite{deheer, brack, kreibig_vollmer}

At the quantum level, linear response theory for an electron gas confined by a positive
jellium background yields a resonance at the Mie frequency $\omega_{\rm M}$ with
a total linewidth
\cite{kubo}
\begin{equation}
\label{gamma_t}
\gamma_{\rm t}(a)=\gamma_{\rm i}+\gamma(a).
\end{equation}
Thus, in addition to the intrinsic linewidth $\gamma_{\rm i}$, we have to
consider a size-dependent contribution which can be expressed as \cite{kubo, barma, yannouleas}
\begin{equation}
\label{gamma_intro}
\gamma(a)=\frac{3v_{\rm F}}{4a}g_0\left(\frac{\varepsilon_{\rm
F}}{\hbar\omega_{\rm M}}\right)
\end{equation}
where $\varepsilon_{\rm F}=\hbar^2k_{\rm F}^2/2m_{\rm e}$ and $v_{\rm F}$ 
are the Fermi energy and velocity, respectively, and $g_0$ is a smooth function 
that will be given in \eqref{g_0}. \cite{yannouleas_85} The size-dependent linewidth 
$\gamma(a)$ results from the decay of the surface plasmon into particle-hole 
pairs by a Landau damping mechanism, which is the dominant decay channel for 
nanoparticle sizes $\unit[0.5]{nm} \lesssim a\lesssim \unit[5]{nm}$ considered in 
this work. For larger clusters, the interaction of the surface plasmon with the external 
electromagnetic field becomes the preponderant source of damping. \cite{kreibig_vollmer}

In a quantum many-body approach, the surface plasmon appears as a collective
excitation of the electron system. Discrete-matrix random phase approximation (RPA)
provides a useful representation since the eigenstates of the correlated
electron system are expressed as superpositions of particle-hole states built
from the Hartree-Fock ground state. Following similar approaches developed for
the study of giant resonances in nuclei, Yannouleas and Broglia
\cite{yannouleas} proposed a partition of the many-body RPA Hilbert space into a
low-energy sector (the restricted subspace), containing particle-hole
excitations with
low energy, and a high-energy sector (the additional subspace). The surface
plasmon arises from a coherent superposition of a large number of basis states of
the restricted subspace. Its energy lies in the high-energy sector, and
therefore the mixture with particle-hole states of the additional subspace
results in the broadening of the collective resonance. 

\begin{figure}[t]
\begin{center}
\includegraphics[width=8truecm]{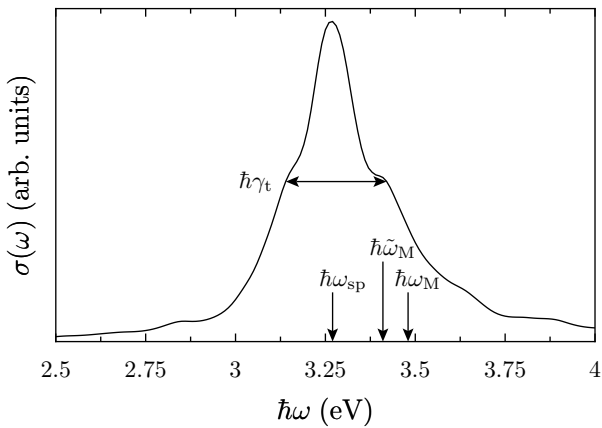}
\caption{\label{fig_absorption} 
Absorption cross section $\sigma(\omega)$ in arbitrary units extracted from TDLDA 
calculations for a sodium cluster containing $N=832$ valence electrons. The 
classical Mie frequency is $\omega_{\rm M}$, while $\tilde\omega_{\rm M}$ is the 
frequency of the surface plasmon resonance taking into account the spill-out effect. 
It has been obtained with the Kohn-Sham self-consistent ground-state density as explained 
in Sec.~\ref{sec_TDLDA}. $\omega_{\rm sp}$ corresponds to the position of the maximum 
of the absorption curve.}
\end{center}
\end{figure}

An alternative to the previous approaches is given by numerical calculations using
the time-dependent local density approximation (TDLDA) within a jellium model.
\cite{ekardt, bertsch_RPA} The absorption cross section 
\begin{equation}
\label{def_cross-section}
\sigma(\omega)=\frac{4\pi e^2\omega}{3c}\sum_f\left|\langle
f|z|0\rangle\right|^2\delta(\hbar\omega-E_f+E_0)
\end{equation}
can be obtained from the dipole matrix elements between the ground state
$|0\rangle$ and the excited states $|f\rangle$ of the electron system with 
energies $E_0$ and $E_f$, respectively. $c$ is the speed of light. In the
absorption cross section, the surface plasmon appears as a broad resonance 
(see Fig.~\ref{fig_absorption}) centered at a frequency $\omega_{\rm sp}$ 
close to $\omega_{\rm M}$ and with a linewidth approximately described by 
\eqref{gamma_t}. 

Finally, the separation in center-of-mass and relative coordinates for the
electron system \cite{gerchikov, weick} within a mean-field approach allows to 
describe the surface plasmon as the oscillation of a collective coordinate, 
which is damped by the interaction with an environment constituted by a large
number of electronic degrees of freedom. Within such a decomposition, the effects of finite 
size, a dielectric material, \cite{weick} or the finite temperature of the electron gas are 
readily incorporated. Since the excitation by a laser field in the optical
range only couples to the electronic center of mass, this approach is
particularly useful.

The experimentally observed resonance frequencies are smaller than the Mie frequency
$\omega_{\rm M}$, \cite{deheer, brechignac92, reiners} and this is qualitatively 
captured by TDLDA calculations since $\omega_{\rm sp}<\omega_{\rm M}$ (see
Fig.~\ref{fig_absorption}). This redshift is usually attributed to the so-called 
{\it spill-out effect}. \cite{brack} The origin of this quantum effect is a 
nonzero probability to find electrons outside the nanoparticle, which results 
in a reduction of the effective frequency for the center-of-mass coordinate.
If a fraction $N_{\rm out}/N$ of the electrons is outside the geometrical 
boundaries of the nanoparticle, we can expect that the electron density within
the nanoparticle is reduced accordingly and the frequency of the surface plasmon 
is given by 
\begin{equation}
\label{omega_M}
\tilde\omega_{\rm M}=\omega_{\rm M}\sqrt{1-\frac{N_{\rm out}}{N}}.
\end{equation}
We will see in Sec.~\ref{sec_coupling} that such an estimation can be formalized
considering the form of the confining potential. It 
has been known for a long time \cite{brack} that $\tilde\omega_{\rm M}>\omega_{\rm sp}$, 
and therefore the spill-out effect is not sufficient to explain the redshift of 
the surface plasmon frequency as it can be seen in Fig.~\ref{fig_absorption}. 
Recently, the coupling to the electronic environment has been invoked as an additional
source of frequency shift. \cite{gerchikov, hagino} In this work we provide an
estimation of such a contribution (analogous to the Lamb shift of atomic physics 
\cite{cohen}) and its parametric dependence on the particle size and electron 
temperature (see Sec.~\ref{sec_delta}). In our approach, we always adopt a
spherical jellium model for the ionic background which is expected to yield
reliable results for not too small nanoparticle sizes. Despite the fact that we have to make
several approximations, we find that this additional shift implies a reduction
of the surface plasmon frequency, in qualitative agreement with the TDLDA
calculations. However it is known \cite{brack} that the experimental resonance
frequency is even lower than the TDLDA prediction. This could be due to the
ionic degrees of freedom.

In femtosecond time-resolved pump-probe experiments on metallic nanoparticles
\cite{bigot, delfatti} the surface plasmon plays a key role. The pump field 
heats the electron system via an excitation of the collective mode. The probe 
then tests the absorption spectrum of the hot electron gas in the nanoparticle, 
and therefore it is important to understand the temperature dependences of the 
surface plasmon linewidth and of the frequency of the resonance.

The paper is organized as follows: In Sec.~\ref{sec_coupling}, we present 
our model and the separation of the electronic degrees of freedom into the 
center of mass and the relative coordinates. The center-of-mass coordinate
provides a natural description of the spatial collective oscillations of the 
electronic cloud around its equilibrium position resulting from a laser 
excitation. Its coupling to the relative coordinates, described within 
the mean-field approximation, is responsible for the broadening of the surface 
plasmon resonance that we evaluate in Sec.~\ref{sec_gamma}. We determine 
the dependence of the surface plasmon linewidth on the size of the nanoparticle 
and explore its low-temperature properties. In Sec.~\ref{sec_spill-out}, we evaluate the 
spill-out effect, focusing on its dependence on size and temperature 
of the nanoparticle. We show by means of TDLDA calculations that the spill-out
effect is not sufficient to describe the redshift of the surface plasmon
frequency as compared to its classical value. In Sec.~\ref{sec_delta}, we propose
an estimation of the environment-induced redshift of the surface plasmon 
resonance which adds to the spill-out effect. Both effects together, spill-out and 
frequency shift due to the electronic environment, could explain the observed 
redshift of the resonance frequency. In Sec.~\ref{sec_pump-probe}, we draw the 
consequences of our findings for pump-probe experiments. The temperature 
dependences of the linewidth and frequency of the surface plasmon resonance peak 
permit to qualitatively explain the time dependence of the measured optical 
transmission as a function of the delay between the pump and the probe laser field. 
We finally conclude and draw the perspectives of this work in Sec.~\ref{sec_ccl}.

%===========================================================================
%===========================================================================
%===========================================================================
%===========================================================================
\section{Coupling of the surface plasmon to electron-hole excitations}
\label{sec_coupling}
Treating the ionic background of the nanoparticle as a jellium sphere of radius 
$a$ with sharp boundaries, the Hamiltonian for the valence electrons is given by
\begin{equation}
\label{H}
H = \sum_{i = 1}^{N} \left[ \frac{p_i^2}{2 m_{\rm e}} + U(r_i) \right] 
+ \frac{e^2}{2} \sum_{\substack{i, j = 1\\(i \neq j)}}^{N}
\frac{1}{\left| {\bf r}_i - {\bf r}_j \right|},
\end{equation}
where ${\bf r}_i$ is the position of the $i^{\rm th}$ particle and 
$r_i=|{\bf r}_i|$. The single-particle confining potential
\begin{equation}
\label{harlomb}
U(r)=\frac{Ne^2}{2a^3}\left(r^2-3a^2\right)\Theta(a-r)-\frac{Ne^2}{r}\Theta(r-a)
\end{equation}
is harmonic with frequency $\omega_{\rm M}=\sqrt{Ne^2/m_{\rm e}a^3}$ inside the 
nanoparticle and Coulomb-like outside. $\Theta(x)$ denotes the Heaviside step 
function. In principle, the photoabsorption cross section \eqref{def_cross-section} 
can be determined from the knowledge of the eigenstates of $H$. However, except for 
clusters containing only few atoms, this procedure is exceedingly difficult, and one has to treat 
this problem using suitable approximation schemes.

%===========================================================================
%===========================================================================
%===========================================================================
\subsection{Separation into collective and relative coordinates}
A particularly useful decomposition \cite{gerchikov} of the Hamiltonian \eqref{H}
can be achieved by introducing the coordinate of the electronic center of mass 
${\bf R}=\sum_i{\bf r}_i/N$ and its conjugated momentum ${\bf P}=\sum_i {\bf p}_i$. 
The relative coordinates are denoted by ${\bf r}_i'={\bf r}_i-{\bf R}$ and 
${\bf p}_i'={\bf p}_i-{\bf P}/N$. Then, the Hamiltonian \eqref{H} can be written as 
\begin{equation}
\label{H_inter}
H=\frac{{\bf P}^2}{2Nm_{\rm e}}+H_{\rm rel}
+\sum_{i=1}^N\left[U(|{\bf r}_i'+{\bf R}|)-U(r_i')\right], 
\end{equation}
where 
\begin{equation}
\label{H_rel_1q}
H_{\rm rel}=\sum_{i = 1}^{N} \left[\frac{{p_i'}^2}{2 m_{\rm e}} + U(r_i') \right] 
+ \frac{e^2}{2} \sum_{\substack{i, j = 1\\(i \neq j)}}^{N}
\frac{1}{\left| {\bf r}_i' - {\bf r}_j' \right|}
\end{equation}
is the Hamiltonian for the relative-coordinate system.

Assuming that the displacement $\bf R$ of the center of mass is small compared
to the size of the nanoparticle, we can expand the last term on the
r.h.s.~of \eqref{H_inter}. To second order, we obtain
\begin{equation}
\label{delta_U}
U(|{\bf r}'+{\bf R}|)-U(r')\simeq
{\bf R}\cdot\boldsymbol{\nabla}U(r')
+\frac 12 \left({\bf R}\cdot\boldsymbol{\nabla}\right)^2 U(r'),
\end{equation}
where the derivatives are taken at ${\bf r}={\bf r}'$ (${\bf R}={\bf 0}$).
Choosing the oscillation axis of the center of mass in the $z$-direction, 
${\bf R}=Z{\bf e}_z$, we obtain 
with \eqref{harlomb}
\begin{equation}
\label{R_1}
{\bf R}\cdot\boldsymbol{\nabla}U(r')=Zm_{\rm e}\omega_{\rm M}^2
\left[z'\Theta(a-r')+\frac{z'a^3}{{r'}^3}\Theta(r'-a)\right]
\end{equation}
and 
\begin{align}
\label{R_2_inter}
\left({\bf R}\cdot\boldsymbol{\nabla}\right)^2 U(r')&=\\
Z^2Ne^2&\left[\frac{1}{a^3}\Theta(a-r')
+\frac{1-3\cos^2{\theta'}}{{r'}^3}\Theta(r'-a)\right].\nonumber
\end{align}
Eq.~\ref{R_1} represents a linear coupling in $Z$ between the center-of-mass and the
relative-coordinate system. In second order, the first term on the r.h.s.~of \eqref{R_2_inter} 
is the dominant contribution to the confinement of the center of mass. The second term on 
the r.h.s.~of \eqref{R_2_inter} is of second order in the coupling and therefore
is neglected compared to the first-order coupling of \eqref{R_1}.

Inserting \eqref{delta_U} and \eqref{R_2_inter} into \eqref{H_inter}, we obtain
\begin{equation}
\label{H_before_decompo}
H=\frac{{\bf P}^2}{2Nm_{\rm e}}+\frac 12\frac{Ne^2}{a^3}{\bf R}^2\sum_{i=1}^N\Theta(a-r_i)
+H_{\rm rel}+H_{\rm c}, 
\end{equation}
where 
\begin{equation}
\label{H_c_1q}
H_{\rm c}=\sum_{i=1}^N{\bf
R}\cdot\left[\boldsymbol{\nabla}U(r_i')\right]\Big|_{{\bf R}={\bf 0}}
\end{equation}
is the coupling between the center-of-mass and the relative coordinates to first 
order in the displacement $\bf R$ of the center of mass according to \eqref{R_1}. 
The remaining sum over $i$ in \eqref{H_before_decompo} yields the number of electrons 
inside the nanoparticle, i.e., $N-N_{\rm out}$ where $N_{\rm out}$ is the number of
spill-out electrons, and finally we rewrite the Hamiltonian as 
\begin{equation}
\label{H_decomposition}
H = H_{\rm cm}+H_{\rm rel}+H_{\rm c}. 
\end{equation}
The Hamiltonian of the center-of-mass system is
\begin{equation}
\label{H_cm_1q}
H_{\rm cm}=\frac{{\bf P}^2}{2Nm_{\rm e}}+\frac 12 Nm_{\rm e}{\tilde
\omega_{\rm M}}^2{\bf R}^2.
\end{equation}
It is the Coulomb tail of the single-particle confinement \eqref{harlomb} which
yields the frequency $\tilde\omega_{\rm M}$ given in \eqref{omega_M} instead of
$\omega_{\rm M}$ for the effective harmonic trap that is experienced by the
center-of-mass system. 

Eq.~\ref{H_decomposition} with \eqref{H_rel_1q}, \eqref{H_c_1q}, and \eqref{H_cm_1q}
recovers up to the second order in $\bf R$ the decomposition derived in 
Ref.~\onlinecite{gerchikov}. However, in contrast to that work, we do not need to appeal 
to an effective potential for the center-of-mass system.

The structure of \eqref{H_decomposition} is typical for quantum dissipative systems: 
\cite{weiss} The system under study ($H_{\rm cm}$) is coupled via $H_{\rm c}$ to an 
environment or ``heat bath" $H_{\rm rel}$, resulting in dissipation and decoherence of the 
collective excitation. In our case the environment is peculiar in the sense that
it is not external to the nanoparticle, but it represents a finite number of
degrees of freedom of the gas of conduction electrons.
If the single-particle confining potential $U$ of 
\eqref{harlomb} were harmonic for all $r$, Kohn's theorem \cite{kohn} would imply 
that the center of mass and the relative coordinates are decoupled, i.e., $H_{\rm c}=0$. 
Thus, for a harmonic potential $U$ the surface plasmon has an infinite lifetime. The Coulomb 
part of $U$ in \eqref{harlomb} leads to the coupling of the center of mass and the 
relative coordinates, and translates into the decay of the surface plasmon. Furthermore, 
it reduces the frequency of the center-of-mass system from $\omega_{\rm M}$ according to
\eqref{omega_M}.

%===========================================================================
%===========================================================================
%===========================================================================
\subsection{Mean-field approximation}
Introducing the usual annihilation operator 
\begin{equation}
\label{b}
b=\sqrt{\frac{Nm_{\rm e}\tilde\omega_{\rm M}}{2\hbar}}Z+\frac{\rm
i}{\sqrt{2Nm_{\rm e}\hbar\tilde\omega_{\rm M}}}P_Z
\end{equation}
where $P_Z$ is the momentum conjugated to $Z$, and the corresponding creation operator 
$b^\dagger$, $H_{\rm cm}$ reads 
\begin{equation}
\label{H_cm}
H_{\rm cm} = \hbar \tilde\omega_{\rm M} b^\dagger b.
\end{equation}

\begin{figure}[t]
\begin{center}
\includegraphics[width=8truecm]{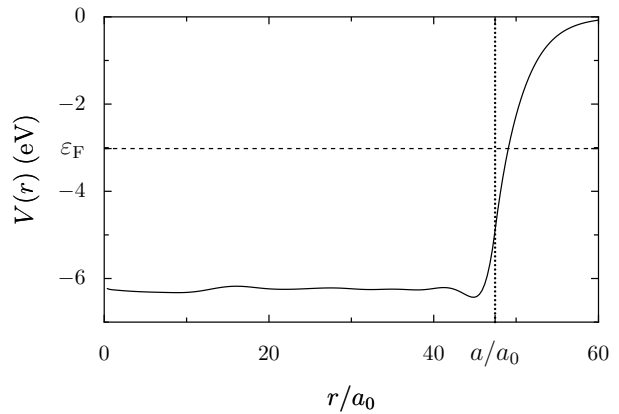}
\caption{\label{fig_potential_fullrange} LDA self-consistent potential $V$ as a function
of the radial coordinate $r$ for a sodium nanoparticle containing
$N=1760$ valence electrons. The radius $a$ is indicated by the vertical dotted 
line. The Fermi level is marked by the dashed line.}
\end{center}
\end{figure}

Since the electron-electron interaction appears in the Hamiltonian
\eqref{H_rel_1q}, it is useful to describe $H_{\rm rel}$ within a mean-field 
approximation. One can write
\begin{equation}
\label{H_rel}
H_{\rm rel} = \sum_\alpha \varepsilon_\alpha c_\alpha^{\dagger} c_\alpha,
\end{equation}
where $\varepsilon_\alpha$ are the eigenenergies in the effective mean-field 
potential $V$ and $c_\alpha^{\dagger}$ ($c_\alpha$) are the creation 
(annihilation) operators associated with the corresponding one-body eigenstates 
$|\alpha\rangle$. The mean-field potential $V$ can be determined with the help
of Kohn-Sham LDA numerical calculations. In Fig.~\ref{fig_potential_fullrange}, we show
it as a function of $r/a_0$ where $a_0\simeq\unit[0.53]{\AA}$ is the Bohr radius, for a
nanoparticle containing $N=1760$ valence electrons. We see that it is relatively
flat at the interior of the nanoparticle, and presents a steep increase at the
boundary. The potential jump is often approximated by a true discontinuity at
$r=a$. However, the details of the self-consistent potential close to the
surface may be crucial for some properties, as we show in Sec.~\ref{sec_TDLDA}.

Inserting \eqref{R_1} into the coupling Hamiltonian $H_{\rm c}$ 
\eqref{H_c_1q}, and expressing the $Z$-coordinate in terms of the
creation and annihilation operators of \eqref{b}, one obtains in second
quantization
\begin{equation}
\label{H_c}
H_{\rm c} =  \Lambda\left(b^\dagger + b\right) \sum_{\alpha\beta} d_{\alpha\beta}
c_\alpha^{\dagger} c_\beta ,
\end{equation}
where 
\begin{equation}
\label{d}
d_{\alpha\beta}=
\langle\alpha|\left[z\Theta(a-r)+\frac{za^3}{r^3}\Theta(r-a)\right]|\beta\rangle
\end{equation}
is the matrix element between two eigenstates of the unperturbed mean-field 
problem. In \eqref{H_c}, we have defined the constant 
$\Lambda=\sqrt{\hbar m_{\rm e}\omega_{\rm M}^3/2N}$ and neglected the 
spill-out for the calculation of the coupling when we expressed $Z$ in terms of
$b$ and $b^\dagger$. The sums over $\alpha$ and $\beta$ in \eqref{H_c} are restricted 
to the additional (high energy) RPA subspace mentioned in the introduction. If initially 
the center of mass is in its first excited state, the 
coupling $H_{\rm c}$ allows for the decay of the collective excitation into 
particle-hole pairs in the electronic environment $H_{\rm rel}$, the so-called
Landau damping. \cite{bertsch, weick, molina}

%===========================================================================
%===========================================================================
%===========================================================================
%===========================================================================
\section{Temperature dependence of the surface plasmon linewidth}
\label{sec_gamma}
In this section, we address a semiclassical calculation of the surface plasmon
linewidth and extend our zero-temperature results of Ref.~\onlinecite{weick} to
the case of finite temperatures. We are interested in nanoparticles with a large
number of confined electrons, typically of the order of $10^2$ to $10^4$. Thus, the
Fermi energy is much larger than the mean one-body level spacing and the 
semiclassical approximation can be applied.

%===========================================================================
%===========================================================================
%===========================================================================
\subsection{Fermi's golden rule}
\label{sec_FGR}
When the displacement of the center of mass is much smaller than the size of the 
nanoparticle, it is possible to linearize the coupling Hamiltonian as it
was done in Sec.~\ref{sec_coupling}. The weak coupling regime allows 
to treat $H_{\rm c}$ as a perturbation to the uncoupled Hamiltonian 
$H_{\rm cm}+H_{\rm rel}$. Assuming that initially the center of mass is in its 
first excited state $|1_{\rm cm}\rangle$, i.e., a surface plasmon is excited, 
two processes limit the surface plasmon lifetime: the decay into the ground state 
with creation of a particle-hole pair and the excitation to the second state 
of the center of mass accompanied by the annihilation of a particle-hole pair. 
Obviously, this last process is only possible at finite temperatures. Then, 
Fermi's golden rule yields the linewidth
\begin{align}
\gamma=\frac{2\pi}{\hbar}
\sum_{\substack{F_{\rm cm}\\ I_{\rm rel}, F_{\rm rel}}}&P_{I_{\rm rel}}
\left|\langle F_{\rm cm}, F_{\rm rel} |H_{\rm c} |1_{\rm cm},I_{\rm rel}
\rangle\right|^2\nonumber\\
&\times 
\delta(\hbar\tilde\omega_{\rm M}-\varepsilon_{F_{\rm cm}}
+\varepsilon_{I_{\rm rel}}-\varepsilon_{F_{\rm rel}})
\end{align}
for the collective state $|1_{\rm cm}\rangle$. In the golden rule, 
$|F_{\rm cm}\rangle$ and $|F_{\rm rel}\rangle$ are the final states of 
center-of-mass and relative coordinates with energy $\varepsilon_{F_{\rm cm}}$ 
and $\varepsilon_{F_{\rm rel}}$, respectively. The probability of finding the 
initial state $|I_{\rm rel}\rangle$ occupied is given in the grand-canonical 
ensemble by the matrix element
\begin{equation}
\label{proba}
P_{I_{\rm rel}}=\frac{\langle I_{\rm rel}|{\rm e}^{-\beta(H_{\rm rel}-\mu
N)}|I_{\rm rel}\rangle}{\Xi}
\end{equation}
of the equilibrium density matrix at the inverse temperature $\beta=1/k_{\rm B}T$, 
$k_{\rm B}$ being the Boltzmann constant. $\mu$ is the chemical potential of the 
electrons in the self-consistent field $V$ and $\Xi$ the grand-canonical partition 
function. Introducing the expression \eqref{H_c} of the coupling Hamiltonian, one finds that 
\begin{equation}
\label{FGR}
\gamma = \Sigma(\tilde\omega_{\rm M})+2\Sigma(-\tilde\omega_{\rm M}),
\end{equation}
where the first term is the rate associated with the 
spontaneous decay of a plasmon, while the second one is the rate for a plasmon 
excitation by the thermal environment. The relative factor of $2$ arises 
from the dipolar transition between the first and second center-of-mass excited states,
$\langle 2_{\rm cm}|b^\dagger|1_{\rm cm}\rangle=\sqrt{2}$. 

In \eqref{FGR}, we have introduced the function
\begin{equation}
\label{gamma_omega}
\Sigma(\omega) = 
\frac{2 \pi}{\hbar^2}
\sum_{\alpha\beta}\left[1-f(\varepsilon_\alpha)\right]f(\varepsilon_\beta)
\left|\Lambda d_{\alpha\beta}\right|^2
\delta\left(\omega-\omega_{\alpha\beta}\right),
\end{equation}
which will be helpful for the evaluation of $\gamma$ and
will also be useful in Sec.~\ref{sec_delta} when the
redshift of the surface plasmon induced by the electronic environment will be
determined. 
In the above expression, $\omega_{\alpha\beta}=(\varepsilon_\alpha-\varepsilon_\beta)/\hbar$ is 
the difference of the eigenenergies in the self-consistent field $V$ and
\begin{equation}
f(\varepsilon)=\frac{1}{{\rm e}^{\beta(\varepsilon-\mu)}+1}
\end{equation}
is the Fermi-Dirac distribution. It is understood that in
\eqref{gamma_omega}, $|\omega_{\alpha\beta}|>\omega_{\rm c}$ where
$\omega_{\rm c}$ is some cutoff separating the restricted subspace that
builds the coherent superposition of the surface plasmon excitation from the
additional subspace at high energies. 
The expression \eqref{gamma_omega} implies the detailed-balance 
relation
\begin{equation}
\label{detbal}
\Sigma(-\omega)={\rm e}^{-\beta\hbar\omega}\Sigma(\omega)
\end{equation}
which allows to write \eqref{FGR} as 
\begin{equation}
\label{gamma}
\gamma=
\Sigma(\tilde\omega_{\rm M}) \left( 1+2{\rm e}^{-\beta\hbar\tilde\omega_{\rm M}}\right).
\end{equation}
This expression shows that an excitation of the surface plasmon to a higher
level is suppressed at low temperatures.

%===========================================================================
%===========================================================================
%===========================================================================
\subsection{Semiclassical low-temperature expansion}
\label{sec_Sigma}
We now use semiclassical techniques to determine the low-temperature behavior of 
$\Sigma(\omega)$. In view of the detailed-balance relation \eqref{detbal} we can
restrict ourselves to positive frequencies $\omega$. In a first step, we need to 
calculate the matrix elements $d_{\alpha\beta}$ defined in \eqref{d}. The 
spherical symmetry of the problem allows to write 
$d_{\alpha\beta}={\cal A}_{l_\alpha l_\beta}^{m_\alpha m_\beta}
{\cal R}(\varepsilon_\alpha,\varepsilon_\beta)$, where an expression for the 
angular part ${\cal A}_{l_\alpha l_\beta}^{m_\alpha m_\beta}$ is given in 
Ref.~\onlinecite{weick}. Dipole selection rules imply $l_\alpha=l_\beta \pm 1$ 
and $m_\alpha=m_\beta$, where $l$ and $m$ are the angular momentum quantum 
numbers. The self-consistent potential is usually approximated by a step-like function, 
$V(r)\simeq V_0 \Theta(r-a)$, where $V_0=\varepsilon_{\rm F}+W$ with $W$ the work 
function of the metal. The accuracy of such an approximation can be estimated by 
comparison with LDA numerical calculations (see Fig.~\ref{fig_potential_fullrange} and 
Ref.~\onlinecite{weick}). 
In the case of strong electronic confinement $V_0\gg\varepsilon_{\rm F}$,
the radial part can be approximated by \cite{yannouleas}
\begin{equation}
\label{radial_ME}
{\cal R}(\varepsilon_\alpha, \varepsilon_\beta)=
\frac{2\hbar^2}{m_{\rm e}a}
\frac{\sqrt{\varepsilon_\alpha\varepsilon_\beta}}
{\left(\varepsilon_\alpha-\varepsilon_\beta\right)^2}.
\end{equation}
The condition of strong confinement implicit in \eqref{radial_ME} assumes that 
the spill-out effect is negligible. Replacing in \eqref{gamma_omega} the sums over 
$\alpha$ and $\beta$ by integrals and introducing the angular-momentum restricted 
density of states (DOS) $\varrho_l(\varepsilon)$, \cite{weick, molina} one gets
\begin{align}
\label{gamma_inter}
\Sigma(\omega)=&\frac{4\pi}{\hbar} 
\int_{\hbar\omega}
^\infty
{\rm d} \varepsilon \left[1-f(\varepsilon)\right]f(\varepsilon-\hbar \omega)
\\
&\times \sum_{\substack{l, m\\l', m'}}
\varrho_{l}(\varepsilon) \varrho_{l'}(\varepsilon- \hbar \omega)
\left[ \Lambda{\cal A}_{ll'}^{mm'}
{\cal R}(\varepsilon, \varepsilon- \hbar \omega) \right]^2, \nonumber 
\end{align}
where a factor of $2$ accounts for the spin degeneracy.

We now appeal to the semiclassical approximation for the two DOS appearing in 
\eqref{gamma_inter} using the Gutzwiller trace formula\cite{gutzwiller} for
the effective radial motion. \cite{weick, molina} The DOS is decomposed 
into a smooth and an oscillating part. With the smooth part 
\begin{equation}
\label{DOS_0}
\varrho_l(\varepsilon)= 
\frac{\sqrt{2 m_{\rm e}a^2 \varepsilon/\hbar^2 -(l+1/2)^2}}{2 \pi \varepsilon}
\end{equation}
of the DOS, and after performing the summation over the angular momentum quantum 
numbers, \eqref{gamma_inter} reads
\begin{equation}
\label{gamma_T_inter}
\Sigma(\omega)=\frac{3 v_{\rm F}}{8a} 
\left(\frac{\hbar \omega_{\rm M}}{\varepsilon_{\rm
F}}\right)^3\frac{\varepsilon_{\rm F}}{\hbar\omega} 
{\cal F}(\mu,\hbar \omega).
\end{equation}
Here, we have introduced
\begin{equation}
\label{F}
{\cal F}(\mu,\hbar \omega)=
\int_{\hbar \omega}^\infty \frac{{\rm d} \varepsilon}{\hbar \omega
}\left[1-f(\varepsilon)\right]f(\varepsilon-\hbar \omega)
H\left(\frac{\varepsilon}{\hbar \omega}\right),
\end{equation}
where
\begin{equation}
\label{H_function}
H(x) = (2x-1) \sqrt{x (x-1)} -\ln{\left(\sqrt{x}+\sqrt{x-1}\right)} 
\end{equation}
is an increasing function. 
The dependence on the chemical potential $\mu$ in \eqref{F} is via the Fermi
functions appearing in this expression. We calculate the function $\cal F$ in
the appendix and this leads for temperatures much smaller than the Fermi
temperature $T_{\rm F}$ to 
\begin{equation}
\label{Sigma}
\Sigma(\omega)=\frac{3 v_{\rm F}}{4 a}\left(\frac{\omega_{\rm M}}{\omega}\right)^3
g\left( \frac{\varepsilon_{\rm F}}{\hbar\omega}, \frac{T}{T_{\rm F}} \right)
\end{equation}
with
\begin{equation}
\label{g_T}
g\left(x, \frac{T}{T_{\rm F}}\right) = g_0(x)+ g_2(x)\left(\frac{T}{T_{\rm F}}\right)^2,
\end{equation}
where \cite{barma, yannouleas}
\begin{align}
\label{g_0}
g_0(x) = \frac{1}{12x^2}&
\Big\{
\sqrt{x(x+1)} \big(4x(x+1)+3\big)  \\
&-3(2x+1) \ln{\left(\sqrt{x}+\sqrt{x+1}\right)} \nonumber\\
&-\Big[ \sqrt{x(x-1)} \big(4x(x-1)+3\big) \nonumber\\
&-3(2x-1) \ln{\left(\sqrt{x}+\sqrt{x-1}\right)}\Big]\Theta(x-1) 
\Big\}\nonumber  
\end{align}
is a monotonically increasing function with $g_0(0)=0$ and $\lim_{x\rightarrow \infty}g_0(x)=1$, 
while 
\begin{align}
\label{g_2}
&g_2(x) = \frac{\pi^2}{24x}
\Big\{
\sqrt{x(x+1)}(6x-1)+\ln{\left(\sqrt{x}+\sqrt{x+1}\right)} \nonumber\\
&-\left[\sqrt{x(x-1)}(6x+1)
+ \ln{\left(\sqrt{x}+\sqrt{x-1}\right)}\right]\Theta(x-1) 
\Big\} 
\end{align}
with $g_2(0)=0$ and $\lim_{x\rightarrow \infty}g_2(x) = \pi^2/6$. For $x$ near
$1$, a nonanalytical $T^{5/2}$-correction coming from \eqref{T_52} has to be added to 
\eqref{g_T}.

Eq.~\ref{gamma} then yields the size- and temperature-dependent surface plasmon
linewidth 
\begin{equation}
\label{gamma_T}
\gamma = \frac{3 v_{\rm F}}{4 a} 
g\left(\frac{\varepsilon_{\rm F}}{\hbar \omega_{\rm M}}, \frac{T}{T_{\rm
F}}\right).
\end{equation}
Note that the exponential factor appearing in \eqref{gamma} is 
irrelevant in our low-temperature expansion. Furthermore, we have replaced 
$\tilde\omega_{\rm M}$ by $\omega_{\rm M}$ for the calculation of $\gamma$. 
As we will see in Sec.~\ref{sec_spill-out}, the spill-out correction to the Mie
frequency scales as $1/a$. Furthermore, \eqref{gamma_T} shows that $\gamma$
scales also as $1/a$. Thus, incorporating the spill-out correction in the
result of \eqref{gamma_T} would yield higher order terms in $1/a$,
inconsistently with our semiclassical expansion that we have restricted to the
leading order.

At $T=0$, we recover with \eqref{gamma_T} the well-known $1/a$ size dependence of the surface plasmon 
linewidth. First found by Kawabata and Kubo, \cite{kubo} this size dependence is 
due to the confinement of the single-particle states in the nanoparticle. 
We also recover the zero-temperature frequency dependence found in Refs.~\onlinecite{barma} 
and \onlinecite{yannouleas}. As a function of $\varepsilon_{\rm F}$, the Landau 
damping linewidth increases linearly for $\varepsilon_{\rm F}\ll\hbar\omega_{\rm M}$ 
and as $\sqrt{\varepsilon_{\rm F}}$ for $\varepsilon_{\rm F}\gg\hbar\omega_{\rm M}$.
The increase of the linewidth is a consequence of the fact that with increasing
Fermi energy the number of relevant particle-hole excitations rises.

Since the function $g_2$ is positive, finite temperatures lead to a broadening 
of the surface plasmon resonance which to leading order is quadratic. As for
$T=0$, the linewidth decreases with increasing size of the nanoparticle like
$1/a$. Based on classical considerations, this has been proposed in Ref.~\onlinecite{yannouleas},
where the authors argued that $\gamma \simeq \bar v/a$ with the average speed
\begin{equation}
\bar v=\frac{3v_{\rm F}}{4}\left[1+\frac{\pi^2}{6}\left(\frac{T}{T_{\rm
F}}\right)^2\right]
\end{equation}
of electrons at the temperature $T$. The result of Ref.~\onlinecite{yannouleas} is 
only relevant in the high energy limit $\varepsilon_{\rm F}\gg\hbar\omega_{\rm
M}$ where it agrees with our general result \eqref{gamma_T}. 

\begin{figure}[t]
\begin{center}
\includegraphics[width=8truecm]{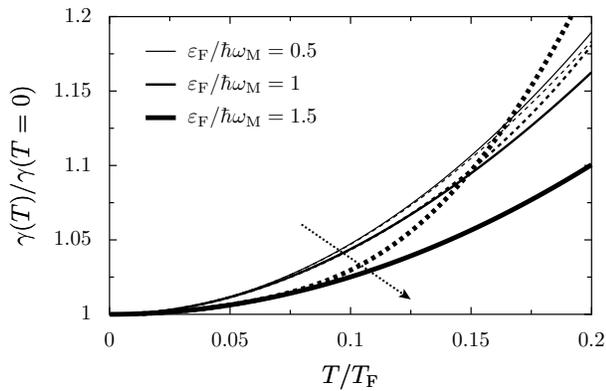}
\caption{\label{fig_gamma}
Surface plasmon linewidth as a function of the temperature, scaled by the
linewidth at $T=0$, for different ratios $\varepsilon_{\rm F}/\hbar \omega_{\rm M}$. 
Solid lines: low-temperature expansion \eqref{gamma_T}. For $\varepsilon_{\rm
F}=\hbar\omega_{\rm M}$, we have accounted for the additional $T^{5/2}$-correction
to $\gamma$, see \eqref{T_52}.
Dashed lines: numerical integration of \eqref{gamma} with 
Eqs.~\ref{gamma_T_inter}--\ref{H_function}.
The arrow indicates the direction of increasing $\varepsilon_{\rm F}/\hbar
\omega_{\rm M}$.
}
\end{center}
\end{figure}

In Fig.~\ref{fig_gamma}, we show the linewidth 
\eqref{gamma_T} as a function of the temperature, scaled by the 
zero-temperature linewidth, for different values of the ratio 
$\varepsilon_{\rm F}/\hbar \omega_{\rm M}$ (solid lines). In order to confirm 
the validity of our low temperature expansion, we compare it to the result 
of a numerical integration (dashed lines) of \eqref{gamma} with 
Eqs.~\ref{gamma_T_inter}--\ref{H_function}. The agreement of the 
expansion \eqref{gamma_T} with 
this direct integration is excellent for low temperatures. For fixed
$T/T_{\rm F}$, the deviation increases with $\varepsilon_{\rm F}/\hbar\omega_{\rm M}$ 
since then, the low-temperature condition $\hbar\omega_{\rm M}\gg k_{\rm B}T$ 
is less and less fulfilled.

There are a number of experiments showing a  broadening of the surface
plasmon linewidth with the temperature. \cite{link, doremus, kreibig} 
Those experimental results on not too small clusters indicate the 
presence of small corrections to the width of the collective surface 
plasmon excitation due to finite temperatures, in agreement with our result
\eqref{gamma_T}.
In Ref.~\onlinecite{link}, absorption measurements 
on gold nanoparticles with a diameter ranging from $9$ to $\unit[25]{nm}$ in aqueous solution 
have shown only a weak temperature effect on the surface plasmon linewidth.
In Ref.~\onlinecite{doremus}, a weak broadening of the plasmon resonance in 
silver and gold nanoparticles of sizes $a=4.25$ to $\unit[10]{nm}$ is reported, accompanied by 
a small redshift of the peak position as the temperature increases. In 
Ref.~\onlinecite{kreibig}, the temperature dependence of small silver clusters (radius 
of $1.6$ to $\unit[10.5]{nm}$) embedded in a glass matrix has been investigated and  
a rather small broadening of the plasmon line has been reported as the temperature increases 
from $1.5$ to $\unit[300]{K}$. 

There are also experiments on very small clusters \cite{haberland, reiners_cpl} showing a
strong temperature effect on the surface plasmon linewidth.
It has been observed in Ref.~\onlinecite{haberland} that the plasmon width of 
mercury clusters increases dramatically with temperature. A systematic study 
of the temperature dependence of the linewidth in small charged sodium nanoparticles 
with $N=8$, $20$, and $40$ valence electrons has been carried out in Ref.~\onlinecite{reiners_cpl}. 
As the temperature of the cluster is increasing, the authors found a pronounced 
broadening of the resonance which goes typically as $\sqrt{T}$. This is in apparent contradiction 
with our result \eqref{gamma_T}. However, for very small particle sizes, an 
additional broadening mechanism becomes important, namely the coupling of the
surface plasmon to quadrupole surface thermal fluctuations. \cite{bertsch, pacheco, tomanek} 

In addition to the smooth part \eqref{DOS_0} of the semiclassical DOS, 
there exists a term which oscillates as a function 
of the energy. The convolution of the two oscillating parts of the DOS in
\eqref{gamma_inter} yields an additional contribution to the plasmon 
linewidth, which oscillates as a function of the size of the nanoparticle. 
This contribution is significant only for very small sizes. 
Its existence has been confirmed by TDLDA calculations. \cite{weick, molina} 
Size-dependent oscillations are also observed in experiments on
very small nanoparticles. \cite{brechignac, rodriguez} 
The oscillations of the linewidth can be expected to be 
smoothed out with increasing temperature because of thermal broadening
suppressing the particle and hole oscillations of the DOS.

%===========================================================================
%===========================================================================
%===========================================================================
%===========================================================================
\section{Spill-out-induced frequency shift}
\label{sec_spill-out}
We now turn to the evaluation of the redshift of the surface plasmon frequency
with respect to the classical Mie value $\omega_{\rm M}$. In this section, we calculate one of
the contributions affecting the resonance frequency, namely the
spill-out effect, while in Sec.~\ref{sec_delta}, we will calculate the
additional frequency shift induced by the electronic environment.

%===========================================================================
%===========================================================================
%===========================================================================
\subsection{Mean-field approximation}
Obtaining the many-body wave function and extracting the number 
of spill-out electrons is very difficult already for very small clusters, 
and practically impossible for larger particles. Therefore we use the mean-field
approximation and treat the electronic degrees of freedom in the 
mean-field one-particle potential $V$ shown in Fig.~\ref{fig_potential_fullrange}. 
Within this approximation, the number of electrons outside of the 
nanoparticle is given by
\begin{equation}
\label{N_out}
N_{\rm out} = 2\int_0^\infty{\rm d} \varepsilon 
\sum_{lm}\varrho_l(\varepsilon) f(\varepsilon) 
\int_{(r>a)}{\rm d}^3{\bf r} \left| \psi_{\varepsilon lm}({\bf r}) \right|^2, 
\end{equation}
where $\varrho_l(\varepsilon)$ is the DOS restricted to a fixed angular
momentum $l$ from \eqref{DOS_0}. The factor of $2$ accounts for the spin degeneracy. 
Because of the spherical symmetry of the problem, the one-particle wave function
\begin{equation}
\psi_{\varepsilon lm}({\bf r}) = \frac{u_{\varepsilon l}(r)}{r} Y_l^m(\theta, \varphi)
\end{equation}
separates into a radial and an angular part given by the spherical harmonics 
$Y_l^m(\theta, \varphi)$. The radial wave functions $u_{\varepsilon l}(r)$
satisfy the reduced Schr\"odinger equation
\begin{equation}
\label{radial_eq}
\left[ -\frac{\hbar^2}{2m_{\rm e}} \frac{{\rm d}^2}{{\rm
d}r^2}+\frac{\hbar^2l(l+1)}{2m_{\rm e}r^2}+V(r) \right]u_{\varepsilon l}(r) = \varepsilon
u_{\varepsilon l}(r)
\end{equation}
with the conditions $u_{\varepsilon l}(0)=0$ and 
$\lim_{r \rightarrow \infty}{\left[u_{\varepsilon l}(r)/r\right]} =0$. Eq.~\ref{radial_eq} 
yields the single-particle eigenenergies $\varepsilon$ in the mean-field potential $V$, 
that we approximate by the step-like potential $V(r)=V_0 \Theta(r-a)$. 

The Fermi function in \eqref{N_out} suppresses contributions to the energy 
integral from values higher than $\varepsilon_{\rm F}$ plus a few $k_{\rm B}T$. 
For low temperatures $k_{\rm B} T \ll W$, the states in the 
continuum do not contribute to $N_{\rm out}$ and we restrict our evaluation to the bound states 
with $\varepsilon<V_0$. \cite{footnote_T}
Defining $k = \sqrt{2m_{\rm e}\varepsilon}/\hbar$ and
$\chi = \sqrt{2m_{\rm e}(V_0-\varepsilon)}/\hbar$, we find for the regular
solutions of \eqref{radial_eq} 
\begin{equation}
u_{\varepsilon l}(r) = \sqrt{\frac{r}{a}}
\begin{cases}
A_{kl}\sqrt{k} J_{l+\frac 12}(kr), &r \leqslant a, \\
B_{kl}\sqrt{\chi} K_{l+\frac 12}(\chi r), &r > a, 
\end{cases}
\end{equation}
where $J_{\nu}(z)$ are Bessel functions of the first kind and $K_{\nu}(z)$
are modified Bessel functions. The normalization constants $A_{kl}$ and $B_{kl}$ are 
given by
\begin{subequations}
\begin{align}
A_{kl} =&\sqrt{\frac{2}{ka C_{kl}}},  \\
B_{kl} =& \sqrt{\frac{2}{\chi a C_{kl}}} \frac{J_{l+\frac
12}(ka)}{K_{l+\frac 12}(\chi a)},
\end{align}
\end{subequations}
with
\begin{align}
C_{kl} = \left[\frac{J_{l+\frac 12}(ka)}{K_{l+\frac 12}(\chi a)}\right]^2&
K_{l-\frac
12}(\chi a)K_{l+\frac 32}(\chi a) \nonumber\\
&-J_{l-\frac 12}(ka)J_{l+\frac 32}(ka).
\end{align}
We therefore obtain for the integral in \eqref{N_out}
\begin{align}
\label{density}
\int_{(r>a)}{\rm d}^3{\bf r} &\left| \psi_{\varepsilon lm}({\bf r}) \right|^2 =
\nonumber \\ 
&\frac{J_{l+\frac 12}^2(ka)}{C_{kl}}
\left[
\frac{K_{l-\frac 12}(\chi a)K_{l+\frac 32}(\chi a)}{K^2_{l+\frac 12}(\chi a)}-1
\right]. 
\end{align}

The summation of this expression over all one-particle states required to obtain $N_{\rm out}$
according to \eqref{N_out} cannot be done exactly. We therefore use a semiclassical 
approximation which provides additional physical insight into the spill-out effect.

%===========================================================================
%===========================================================================
%===========================================================================
\subsection{Semiclassical low-temperature expansion of the number of spill-out electrons}
\label{sec_spill-out_sc}
The integral of the electronic density \eqref{density} increases with the energy. 
Combined with the increasing DOS in \eqref{N_out} and the Fermi function providing 
an energy cutoff, this allows to conclude that the spill-out is dominated by the 
energies near the Fermi level. In addition, in most of the metallic 
nanoparticles, we have $\varepsilon_{\rm F} \sim W \gg \Delta$, where
$\Delta$ is the mean single-particle level spacing. The semiclassical limit, in which 
$ka$ and $\chi a$ must be much larger than one, 
then applies. In this limit, we obtain for the integral \eqref{density} of the density outside 
of the nanoparticle 
\begin{equation}
\label{density_sc}
\int_{(r>a)}{\rm d}^3{\bf r} \left| \psi_{\varepsilon lm}({\bf r}) \right|^2 \simeq 
\frac{\varepsilon}{\chi a V_0},
\end{equation}
to first order in $1/ka$ and $1/\chi a$.
Since this result does not depend on the angular momentum quantum numbers 
$l$ and $m$, the total DOS $\sum_{lm}\varrho_l(\varepsilon)$ (whose explicit expression
can be found in Ref.~\onlinecite{weick}) is sufficient to determine the
number of spill-out electrons. Inserting \eqref{density_sc} into \eqref{N_out} then yields
\begin{equation}
\label{N_out_inter}
N_{\rm out}\simeq\frac{4m_{\rm e}a^2}{3\pi\hbar^2V_0}
\int_0^{V_0} {\rm d}\varepsilon\,f(\varepsilon)\frac{\varepsilon^{3/2}}{\sqrt{V_0-\varepsilon}}. 
\end{equation}
Applying the low-temperature Sommerfeld expansion, \cite{ashcroft} we obtain
\begin{equation}
\label{N_out_T}
N_{\rm out}=\frac{\left(k_{\rm F}a\right)^2}{6\pi}
\zeta\left(\frac{\varepsilon_{\rm F}}{V_0}, \frac{T}{T_{\rm F}}\right) 
\end{equation}
with
\begin{equation}
\label{zeta}
\zeta\left(x, \frac{T}{T_{\rm F}}\right)=
\zeta_0(x)+\zeta_2(x)\left(\frac{T}{T_{\rm F}}\right)^2,
\end{equation}
where 
\begin{equation}
\zeta_0(x) = \frac {1}{x} \left[-\sqrt{x(1-x)}(2x+3)
+ 3\arcsin{\sqrt{x}}\right]
\end{equation}
and 
\begin{equation}
\zeta_2(x)=\frac{\pi^2}{3}  \left(\frac{x}{1-x}\right)^{3/2}(2-x).
\end{equation}
Note that the upper bound of the integral over the energy in
\eqref{N_out_inter} has been replaced by $V_0$ since we neglect 
exponentially suppressed contributions from higher energies. Therefore
our Sommerfeld expansion is reliable for temperatures  
$T/T_{\rm F}\lesssim 1-\varepsilon_{\rm F}/V_0$.

At zero temperature the number of spill-out 
electrons increases smoothly with increasing Fermi energy, reaching its maximal 
value $(k_{\rm F}a)^2/4$ at $\varepsilon_{\rm F} = V_0$. According to 
\eqref{N_out_T} and \eqref{zeta}, the number of spill-out electrons increases 
with temperature. This is expected since the evanescent part 
of the wave function increases with the energy of the occupied states.

Scaling the result \eqref{N_out_T} with the total number of electrons in
the nanoparticle, $N = 4(k_{\rm F}a)^3/9\pi$, we obtain
\begin{equation}
\label{N_out_T_over_N}
\frac{N_{\rm out}}{N} = \frac{3}{8k_{\rm F}a}
\zeta\left(\frac{\varepsilon_{\rm F}}{V_0}, \frac{T}{T_{\rm F}}\right). 
\end{equation}
This relative spill-out scales to first order as $1/a$, and is therefore 
negligible for large particles. As the work function $W$ depends on the size of
the nanoparticle like $W = W_\infty+\alpha/a$, \cite{seidl} the number of spill-out electrons
\eqref{N_out_T_over_N} acquires, via the dependence on $V_0$, higher-order
corrections to the $1/a$ scaling. We neglect these terms because they are of the
same order as terms neglected in our semiclassical expansion. Therefore we 
approximate $W$ by its bulk value $W_\infty$. The dependence of
\eqref{N_out_T_over_N} on $a$ can be interpreted by observing that the spill-out is a 
surface effect so that $N_{\rm out}$ increases only with $a^2$.
Inserting \eqref{N_out_T_over_N} 
into \eqref{omega_M}, one can calculate the spill-out-induced redshift of the 
surface plasmon resonance. 
This redshift increases for decreasing sizes and for increasing temperatures,
in qualitative agreement with experiments. \cite{brechignac, doremus, kreibig}

\begin{figure}[t]
\begin{center}
\includegraphics[width=8truecm]{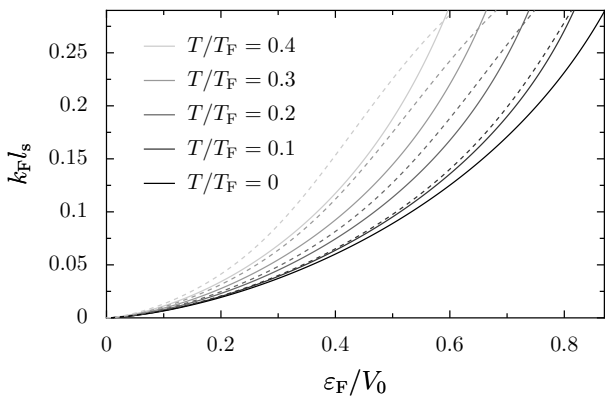}
\caption{\label{kF_ls} 
$k_{\rm F} l_{\rm s}$ from \eqref{ls_T} as
a function of $\varepsilon_{\rm F}/V_0$ at zero temperature (black line) and 
for finite temperatures (solid gray lines). The dashed lines result from 
the numerical integration of \eqref{N_out_inter}. In the figure, 
$T/T_{\rm F}$ increases from bottom to top.}
\end{center}
\end{figure}

One can define a spill-out length as the depth
\begin{equation}
\label{def_ls}
l_{\rm s} = \frac 13 \frac{N_{\rm out}}{N}a 
\end{equation}
of the spill-out layer. Inserting \eqref{N_out_T_over_N} into \eqref{def_ls} yields 
the size-independent spill-out length
\begin{equation}
\label{ls_T}
k_{\rm F} l_{\rm s} = \frac 18
\zeta\left(\frac{\varepsilon_{\rm F}}{V_0}, \frac{T}{T_{\rm F}}\right), 
\end{equation}
which is represented in Fig.~\ref{kF_ls} as a function of the ratio
$\varepsilon_{\rm F}/V_0$ for different temperatures. The result is compared 
with the results of a numerical integration of \eqref{N_out_inter} (dashed
lines). This confirms our expectations about the validity of our result, 
namely for low temperatures and for a reasonable ratio $\varepsilon_{\rm F}/V_0$.

%===========================================================================
%===========================================================================
%===========================================================================
\subsection{Number of spill-out electrons: Semiclassics vs.~LDA}
\label{sec_TDLDA}
\begin{figure}[t]
\begin{center}
\includegraphics[width=8truecm]{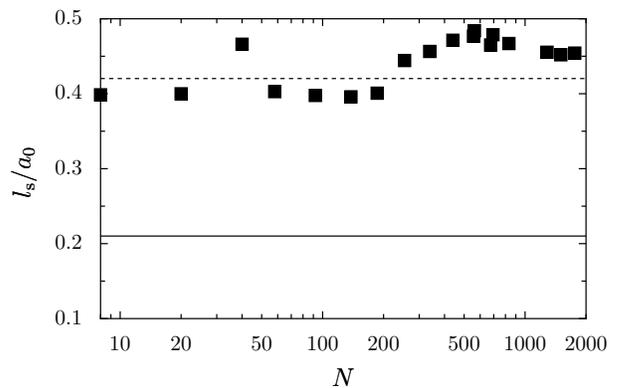}
\caption{\label{fig_l_so} Spill-out length $l_{\rm s}$ at $T=0$ in units of the 
Bohr radius $a_0$ as a function of the total number of electrons $N$, after
\eqref{ls_T}  (solid line) for sodium nanoparticles. 
The dashed line is twice the result of \eqref{ls_T}, and is obtained considering 
the effective radius $a_{\rm eff}$ for the approximated self-consistent potential, 
namely $V(r)=V_0 \Theta(r-a_{\rm eff})$. The squares result from LDA
calculations.}
\end{center}
\end{figure}

In this section, we compare our semiclassical evaluation of the spill-out effect
at zero temperature with LDA calculations on spherical sodium nanoparticles. 
One possible way to 
estimate $N_{\rm out}$ is to use the LDA, which allows to compute the 
spherically symmetric self-consistent electronic ground-state density for 
nanoparticles with closed electronic shells. Integrating the density outside the 
nanoparticle then yields an approximation to $N_{\rm out}$, and thus to 
$l_{\rm s}$ according to \eqref{def_ls}. An estimation of the spill-out 
length from \eqref{ls_T} by means of our semiclassical theory at zero 
temperature gives $l_{\rm s} \simeq 0.2\, a_0$ for sodium nanoparticles, 
while Madjet and collaborators obtained on the basis of Kohn-Sham and 
Hartree-Fock calculations $l_{\rm s}$ around $0.55\, a_0$ for clusters of 
size $N=8$--$196$. \cite{madjet} With our LDA calculations, we obtain 
$l_{\rm s}$ of the order of $0.45\,a_0$ for all closed-shell sizes between $N=8$ and  $N=1760$ 
(see the squares in Fig.~\ref{fig_l_so}).

\begin{figure}[t]
\begin{center}
\includegraphics[width=8truecm]{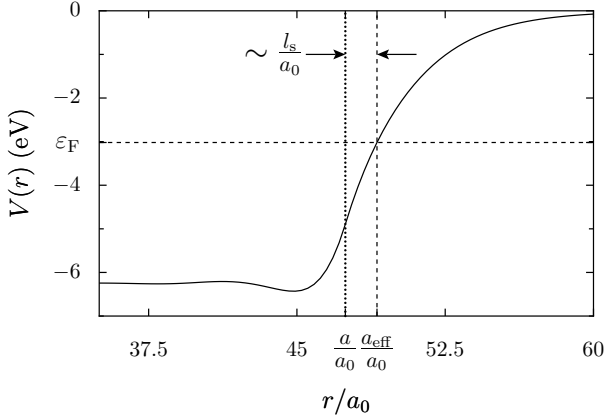}
\caption{\label{fig_potential} Detail of the LDA self-consistent potential $V$
close to the boundary of the cluster as a function
of the radial coordinate $r$ for a sodium nanoparticle containing
$N=1760$ valence electrons. The radius $a$ is indicated by the vertical dotted 
line, and the effective radius defined by
$V(a_{\rm eff}\simeq a+l_{\rm s})=\varepsilon_{\rm F}$ is indicated by the 
dashed line. The Fermi level corresponds to the horizontal dashed line.}
\end{center}
\end{figure}

The fact that the semiclassical spill-out length is significantly smaller than
that of LDA is a consequence of our assumption of a step-like potential
for $V$. Indeed, the LDA self-consistent potential $V$ shown in Fig.~\ref{fig_potential} deviates 
from the form $V(r)=V_0 \Theta(r-a)$ that we have used. As one can see in 
Fig.~\ref{fig_potential}, the Fermi level does not coincide with $V(a)$. Defining 
the effective radius of the nanoparticle for the spill-out effect by means of 
$V(a_{\rm eff})=\varepsilon_{\rm F}$, it seems appropriate to approximate the 
self-consistent potential by $V(r)=V_0 \Theta(r-a_{\rm eff})$. An estimation from 
our LDA calculations gives $a_{\rm eff} \simeq a+l_{\rm s}$ for all sizes between 
$N=8$ and $N=1760$. Using this effective radius in \eqref{N_out_T}
does not change our results for $N_{\rm out}$ and $l_{\rm s}$ since $a\gg l_{\rm s}$.
However, the spill-out length is defined from the
geometrical radius $a$ of the ionic jellium background of the nanoparticle.
Since $a_{\rm eff}\simeq a+l_{\rm s}$, it actually yields an effective spill-out
length $l_{\rm s}^{\rm eff}\approx2l_{\rm s}$ and approximately doubles our result for $N_{\rm out}$.
The improved $T=0$ result for the spill-out length is
represented by the dashed line in Fig.~\ref{fig_l_so}. It yields good
agreement with the spill-out length as deduced from the LDA calculations 
(squares in Fig.~\ref{fig_l_so}). Note that there is no need to consider an effective 
radius of the nanoparticle in the calculation of the resonance width 
presented in Sec.~\ref{sec_gamma}. Indeed, the linewidth $\gamma$ scales as $1/a$. 
Replacing $a$ by an effective radius for the linewidth $\gamma$ would thus only lead 
to higher-order corrections.

%===========================================================================
%===========================================================================
%===========================================================================
\subsection{Redshift of the surface plasmon resonance}
\begin{figure}[t]
\begin{center}
\includegraphics[width=8truecm]{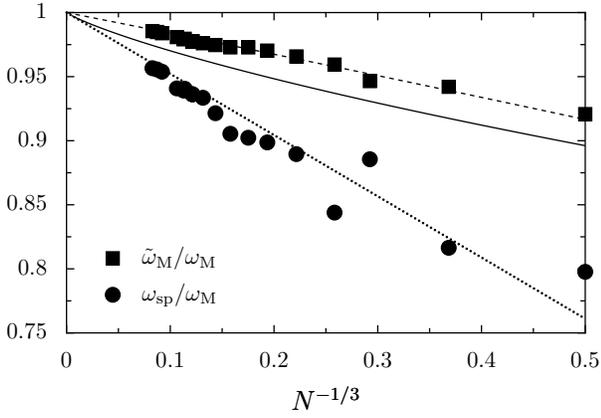}
\caption{\label{fig_TDLDA} Surface plasmon frequency as a function of
the number $N$ of valence electrons in the nanoparticle. 
Squares: 
$\tilde\omega_{\rm M}/\omega_{\rm M}$, 
frequencies deduced from the LDA number of spill-out electrons
according to \eqref{omega_M}. 
Dashed line:  
semiclassical evaluation of \eqref{N_out_T_over_N} with \eqref{omega_M}. 
Dots: 
$\omega_{\rm sp}/\omega_{\rm M}$, 
frequencies obtained by fitting the TDLDA absorption curves with a
Mie-type cross section. \cite{deheer}
Dotted line: linear fit to the dots.
Solid line: 
sum of the spill-out effect (dashed line) and the
environment-induced redshift of the resonance \eqref{shifts}.
All results shown here are at zero temperature for the case of sodium. 
The points shown represent all closed shell sizes between $N=8$ and $N=1760$.}
\end{center}
\end{figure}

We now examine the redshift of the surface plasmon resonance by means of 
TDLDA calculations. In Fig.~\ref{fig_TDLDA}, we show the frequencies deduced
from the LDA number of spill-out electrons for various closed-shell 
nanoparticle sizes between $N=8$ and $N=1760$, where $N_{\rm out}$ is
incorporated according to \eqref{omega_M} (squares). The dashed 
line is our semiclassical result from \eqref{N_out_T_over_N} with
\eqref{omega_M}, where we have taken into account the effective radius for
the self-consistent potential, as discussed in the preceding section. We see
that our analytical expression for the spill-out is in a good agreement with the
LDA calculations, and that the redshift is increasing with decreasing size as
predicted by \eqref{N_out_T_over_N}. 

Alternatively the resonance frequency can be extracted directly from the 
absorption cross section $\sigma(\omega)$ defined in \eqref{def_cross-section}
and calculated from the TDLDA response function (see Fig.~\ref{fig_absorption}). 
An upper bound for the resonance energy is given by the root mean square \cite{brack, bohigas}
\begin{equation}
\label{sum_rule}
\sqrt{\langle\omega^2\rangle}=
\sqrt{\frac{\int_0^\infty{\rm d}\omega \ \omega^2 \sigma(\omega)}
{\int_0^\infty {\rm d} \omega \ \sigma(\omega)}},
\end{equation}
providing a lower bound for the redshift of the surface plasmon frequency from the 
Mie value. In the spherical jellium model, the frequency deduced from 
\eqref{sum_rule} coincides with $\tilde\omega_{\rm M}$, \cite{brack, bertsch_ekardt} 
the Mie frequency redshifted by the spill-out effect \eqref{omega_M}. We 
have checked with our TDLDA calculations that the frequencies deduced from 
$N_{\rm out}$, i.e., from the electronic ground-state self-consistent density
correspond to the frequencies obtained from \eqref{sum_rule}, up to the
numerical error. At zero temperature, the TDLDA thus fulfills the sum rule
\cite{brack} $\sqrt{\langle\omega^2\rangle}=\tilde\omega_{\rm M}$. 

The dots in Fig.~\ref{fig_TDLDA} represent the maxima $\omega_{\rm sp}$ of the 
absorption curves obtained by fitting the TDLDA results with a 
Mie-type cross section. \cite{deheer} It is clear from
Fig.~\ref{fig_TDLDA} that the frequency obtained by \eqref{omega_M} from 
the number of spill-out electrons (squares) overestimates the resonance 
frequency $\omega_{\rm sp}$ (see also Fig.~\ref{fig_absorption}). \cite{HF} It 
has been noticed in Ref.~\onlinecite{reiners} that the spill-out effect is not
sufficient to describe the experimentally observed resonance frequency.

In the following section, we propose to interpret the discrepancy between
$\omega_{\rm sp}$ given by the 
TDLDA and $\tilde\omega_{\rm M}$ deduced from the spill-out by means of
the coupling of the surface plasmon mode to electron-hole excitations. This
coupling results in a shift of the surface plasmon frequency which adds to
the effect of the spill-out.

%===========================================================================
%===========================================================================
%===========================================================================
%===========================================================================
\section{Environment-induced frequency shift}
\label{sec_delta}
In the absence of the coupling $H_{\rm c}$, the energy of the $n^{\rm th}$ 
eigenstate of the center-of-mass system is given by the eigenenergies 
${\cal E}^{(0)}_n=n\hbar\tilde\omega_{\rm M}$ of $H_{\rm cm}$ given in
\eqref{H_cm}. However, the coupling Hamiltonian \eqref{H_c} 
perturbs the eigenstates of $H_{\rm cm}$. 
The leading contribution to the 
resulting shift of the eigenenergies ${\cal E}^{(0)}_n$ is determined using 
perturbation theory in $H_{\rm c}$. While the first order does not 
contribute due to the selection rules contained in the coupling \eqref{H_c}, 
there are four second-order processes involving virtual particle-hole pairs. 
In addition to the two resonant processes mentioned in Sec.~\ref{sec_FGR}, i.e.,  
decay into the ground state with creation of a particle-hole pair and 
excitation to a higher state accompanied by the annihilation of a particle-hole
pair, there exist also two antiresonant processes.
A plasmon can be excited to a higher collective 
state by creating a particle-hole pair, or a plasmon 
can decay by destroying a particle-hole pair. 
Taking into account all four processes, we obtain the 
resonance energy to second order in the coupling
\begin{equation}
\label{omega_sp}
\hbar\omega_{\rm sp}={\cal E}^{(2)}_1-{\cal
E}^{(2)}_0=\hbar\left(\tilde\omega_{\rm M}-\delta\right)
\end{equation}
with the frequency shift 
\begin{align}
\delta=\frac{1}{\hbar}
\sum_{\substack{F_{\rm cm}\\ I_{\rm rel}, F_{\rm rel}}}P_{I_{\rm rel}}
&\left(
\frac{\left|\langle F_{\rm cm},F_{\rm rel}|H_{\rm c}|1_{\rm cm},I_{\rm
rel}\rangle\right|^2}
{\varepsilon_{F_{\rm cm}}-\hbar\tilde\omega_{\rm M}
+\varepsilon_{F_{\rm rel}}-\varepsilon_{I_{\rm rel}}}
\right.\nonumber\\
&\left.
-\frac{\left|\langle F_{\rm cm},F_{\rm rel}|H_{\rm c}|0_{\rm cm},I_{\rm
rel}\rangle\right|^2}
{\varepsilon_{F_{\rm cm}}+\varepsilon_{F_{\rm rel}}-\varepsilon_{I_{\rm rel}}}
\right),
\end{align}   
where we use the same notations as in Sec.~\ref{sec_FGR}. Writing explicitly the
coupling Hamiltonian $H_{\rm c}$ \eqref{H_c} in the preceding expression, we obtain 
\begin{equation}
\label{delta}
\delta = \frac{2}{\hbar^2} {\cal P} \sum_{\alpha\beta} 
\left[1-f(\varepsilon_\alpha)\right] f(\varepsilon_\beta)\left|\Lambda
d_{\alpha\beta}\right|^2
\frac{\omega_{\alpha\beta}}{\omega_{\alpha\beta}^2-\tilde\omega_{\rm M}^2}.
\end{equation}
Here, $\cal P$ denotes the Cauchy principal value. 
The resonance frequency $\omega_{\rm sp}$ \eqref{omega_sp} thus contains 
a correction with respect to the value induced by the spill-out effect. As we 
will show, the shift $\delta$ is positive, and thus redshifts the plasmon 
resonance from $\tilde\omega_{\rm M}$.

According to \eqref{gamma_omega} and \eqref{delta}, the energy shift $\delta$ is 
related to the function $\Sigma(\omega)$ through the Kramers-Kronig relation
\begin{equation}
\label{KK}
\delta=\frac{1}{\pi}{\cal P}\int_{-\infty}^{+\infty} {\rm d}\omega
\frac{\omega \Sigma(\omega)}{\omega^2-\tilde\omega_{\rm M}^2}.
\end{equation}
In \eqref{KK}, the frequency $\tilde\omega_{\rm M}$ appearing in the
denominator can be replaced by $\omega_{\rm M}$. Indeed, the function
$\Sigma(\omega)$ of \eqref{Sigma} is proportional to $1/k_{\rm F}a$. Thus, taking into
account the spill-out effect in the evaluation of $\delta$ would yield higher
order terms in powers of $1/k_{\rm F}a$, that we neglect in the semiclassical
limit. Furthermore, we have to restrict the integral over the frequency $\omega$
by introducing the cutoff $\omega_{\rm c}$ discussed in Sec.~\ref{sec_FGR}.
It arises from the fact that the particle-hole pairs that participate to $\delta$
in \eqref{delta} belong to the high-energy sector of the RPA Hilbert
space, while the surface plasmon excitation is the superposition of
particle-hole pairs of the restricted low-energy subspace. The TDLDA absorption
cross section shows a large excitation peak at the frequency $\omega_{\rm sp}$
which supports almost all of the dipole strength. This peak is surrounded by
particle-hole excitations that become noticeable for frequencies larger than
$\sim\omega_{\rm M}-\eta\gamma$, where $\eta$ is a constant of the order of unity. 
Thus for the purpose of calculating the integral \eqref{KK} we can take the
cutoff at $\omega_{\rm M}-\eta\gamma$ and approximate  
\begin{equation}
\label{delta_inter}
\delta\simeq\frac{1}{\pi}{\cal P}\int_{\omega_{\rm M}-\eta\gamma}^{+\infty} {\rm d}\omega
\frac{\omega \Sigma(\omega)}{\omega^2-\omega_{\rm M}^2}.
\end{equation}

For frequencies $\omega$ larger than $\omega_{\rm M}-\eta\gamma$, the function
$g$ appearing in \eqref{Sigma} can be replaced by its asymptotic expansion for 
$\varepsilon_{\rm F}\ll\hbar\omega$,
\begin{equation}
g\left(x, \frac{T}{T_{\rm F}}\right)\simeq
\left[\frac{8}{15}+\frac{2\pi^2}{9}\left(\frac{T}{T_{\rm F}}\right)^2\right]\sqrt{x}.
\end{equation}
Inserting this expression into \eqref{delta_inter}, and 
performing the remaining integral over $\omega$ in the semiclassical limit $k_{\rm F}a\gg1$,
we arrive at 
\begin{align}
\delta\simeq\frac{3v_{\rm F}}{4a}
&\sqrt{\frac{\varepsilon_{\rm F}}{\hbar\omega_{\rm M}}}
\left[\ln{\left(\frac{4\omega_{\rm M}}{\eta\gamma}\right)}
-\frac{\pi}{2}-\frac 43\right]\nonumber\\
&\times\left[\frac{4}{15\pi}+\frac{\pi}{9}\left(\frac{T}{T_{\rm
F}}\right)^2\right].
\end{align}
We remark that the dependence of this result on the cutoff is only
logarithmic.

Inserting our expression \eqref{gamma_T} for the linewidth $\gamma$,
we obtain to second order in $T/T_{\rm F}$
\begin{equation}
\label{shifts}
\delta = \frac{3 v_{\rm F}}{4 a} 
j \left( \frac{\varepsilon_{\rm F}}{\hbar \omega_{\rm M}}
, \frac{T}{T_{\rm F}}\right)
\end{equation}
where 
\begin{equation}
j\left(x, \frac{T}{T_{\rm F}}\right)=j_0(x)+j_2(x)\left(\frac{T}{T_{\rm F}}\right)^2, 
\end{equation}
with 
\begin{equation}
j_0(x)=\frac{4\sqrt{x}}{15\pi}
\left[\ln{\left(\frac{8k_{\rm F}a}{3\eta xg_0(x)}\right)}-\frac{\pi}{2}-\frac 43\right]
\end{equation}
and 
\begin{align}
j_2(x)=\frac{\pi\sqrt{x}}{9}
\left[\ln{\left(\frac{8k_{\rm F}a}{3\eta xg_0(x)}\right)}-\frac{\pi}{2}-\frac 43
-\frac{12}{5\pi^2}\frac{g_2(x)}{g_0(x)}\right].
\end{align}
The functions $g_0$ and $g_2$ are defined in \eqref{g_0} and \eqref{g_2},
respectively. While the linewidth \eqref{gamma_T} goes as $1/a$, the frequency
shift scales as $1/a$ up to a logarithmic factor.
This redshift increases with temperature and adds to the
redshift arising from the spill-out effect discussed in Sec.~\ref{sec_spill-out}.

The importance of the shift $\delta$ can be seen in Fig.~\ref{fig_TDLDA}.
There, the position of the surface plasmon resonance peak for sodium clusters 
calculated from TDLDA (dots) is in qualitative agreement with our 
semiclassical result (solid line) taking into account the shift $\delta$ and 
the spill-out (see Eqs.~\ref{omega_M} and \ref{N_out_T_over_N} in 
Sec.~\ref{sec_spill-out}). \cite{footnote_TDLDA} In Fig.~\ref{fig_TDLDA}, we have used $\eta=1/2$ 
for the frequency cutoff in \eqref{delta_inter}. Our approximate expression 
for $\delta$ does not allow us to obtain a quantitative agreement with the 
position $\omega_{\rm sp}$ of the resonance frequency shown by the dots. 
This is not surprising considering the approximations needed in order to derive 
an analytical result.
First, the expression for $\delta$ is based on our result \eqref{Sigma}
for $\Sigma(\omega)$ which was derived under the assumption of perfectly
confined electronic states. Thus, the delocalized self-consistent single-particle
states have not been accurately treated in the calculation of $\delta$. 
Second, the cutoff introduced above is a rough estimate of the energy beyond
which particle-hole excitations couple to the surface plasmon. Despite those 
approximations, our estimate implies an increase of the redshift beyond that caused by 
the spill-out (dashed line in Fig.~\ref{fig_TDLDA}).
Comparing the two effects leading to a redshift of the surface plasmon frequency, we 
find that they have the same size and temperature dependence, and are of the same 
order of magnitude. Therefore, one has to take into account both contributions in 
quantitative descriptions of the surface plasmon frequency. 

For zero temperature, the shift $\delta$ has also been considered in Refs.~\onlinecite{gerchikov} and 
\onlinecite{hagino}. The authors of Ref.~\onlinecite{gerchikov} have used the 
separation of the collective center-of-mass motion from the relative
coordinates. Their coupling between the two subsystems which is only
nonvanishing outside the nanoparticle leads to a 
shift that they have numerically evaluated by means of the RPA plus
exchange in the case of small charged sodium clusters. In Ref.~\onlinecite{hagino},
the authors assumed a certain expression for the coupling, and a variational RPA 
calculation was used to obtain an analytical expression of the environment-induced redshift. 
In contrast to our findings, Refs.~\onlinecite{gerchikov} and \onlinecite{hagino} 
obtained a shift $\delta$ proportional to the number of spill-out electrons. 

For very small clusters ($N$ between $8$ and $92$), a nonmonotonic behavior of the 
resonance frequency as a function of the size of the nanoparticle has been observed 
experimentally. \cite{reiners} This behavior is in qualitative agreement with our 
numerical calculations (see dots in Fig.~\ref{fig_TDLDA}) and 
can be understood in the following way. We have shown in 
Refs.~\onlinecite{weick} and \onlinecite{molina} that the linewidth of the 
surface plasmon resonance presents oscillations as a function of the size of 
the nanoparticle. Furthermore, we have shown in this section that the 
linewidth and the environment-induced shift are related through the Kramers-Kronig
transform \eqref{KK}. Thus, the shift $\delta$ should also present
oscillations as a function of the size of the cluster. This is in contrast to 
the spill-out effect discussed in Sec.~\ref{sec_spill-out} where the oscillating 
character of the DOS leads to a vanishing contribution, as confirmed by the
LDA calculations (see squares in Fig.~\ref{fig_TDLDA}). Those
significantly different behaviors could permit to distinguish between the two
mechanisms contributing to the redshift of the surface plasmon frequency with respect to
the classical Mie value.

Using temperature-dependent TDLDA calculations, Hervieux and Bigot \cite{hervieux} 
have recently found a nonmonotonic behavior of the energy shift of the 
surface plasmon frequency as a function of the temperature for a given
nanoparticle size. They observed a redshift of the surface plasmon resonance up to a certain 
critical temperature (e.g., $\unit[1000]{K}$ and $\unit[2500]{K}$ for Na$_{138}$ and Na$_{139}^+$, 
respectively), followed by a blueshift of the resonance at higher temperatures.
This behavior is not present in our theory. The authors of
Ref.~\onlinecite{hervieux} attribute the nonmonotonic temperature dependence to the coupling of 
the surface plasmon to bulklike extended states in the continuum, which causes a blueshift, 
as one can expect for a bulk metal. However, we have restricted ourselves
to low temperatures compared to the work function of the nanoparticle, where we can
neglect those extended states in the evaluation of the spill-out. The 
critical temperature of a thousand degrees is much smaller than the Fermi temperature 
for metals, such that our treatment should remain a good approximation.

%===========================================================================
%===========================================================================
%===========================================================================
%===========================================================================
\section{Time evolution of the optical transmission in a pump-probe configuration}
\label{sec_pump-probe}
We now examine the experimental consequences of the temperature dependence of
the Landau damping linewidth \eqref{gamma_T} and of the energy shifts induced by 
the spill-out (Eqs.~\ref{omega_M} and \ref{N_out_T_over_N}) and by the
electronic environment \eqref{shifts}.

We focus on the absorption cross section defined in 
\eqref{def_cross-section}. Assuming it to be of the Breit-Wigner form, we have
\begin{equation}
\label{Breit-Wigner}
\sigma(\omega, T) = s(a)
\frac{\gamma(T)/2}{\left[ \omega- \omega_{\rm sp}(T) \right]^2+\left[ \gamma(T)/2 \right]^2},
\end{equation}
where $\omega_{\rm sp}(T)$ is the temperature-dependent resonance frequency of 
the surface plasmon excitation given in \eqref{omega_sp}.
$s(a)$ is a size-dependent normalization prefactor. For the spill-out, 
we consider the effective radius discussed in Sec.~\ref{sec_TDLDA} which doubles the results of 
\eqref{N_out_T_over_N}.

Unfortunately, to the best of our knowledge, systematic experimental investigations of 
the shape of the absorption cross section as a function of temperature are not
available. However, an indirect approach is offered by pump-probe experiments, 
\cite{bigot, delfatti} where the nanoparticles are excited by an intense laser 
pulse. After a given time delay, the system is probed with a weak laser field, 
measuring the transmission ${\cal T}$ of the nanoparticles. After the excitation of 
a surface plasmon, its energy is transferred on the femtosecond timescale to the 
electronic environment, resulting in the heating of the latter. On a much longer
timescale of typically a picosecond, the equilibration of the electrons with the phonon 
heat bath results in the decrease of the temperature of the electronic system with time. 
Such a process is not included in our theory.

An experimentally accessible quantity \cite{bigot, delfatti} is the differential 
transmission $\Delta{\cal T}/{\cal T}=({\cal T}_{\rm on}-{\cal T}_{\rm off})/{\cal T}_{\rm
off}$, i.e., the normalized difference of transmissions with and without the
pump laser field. It is related to the absorption cross section by means of 
\begin{equation}
\label{transmission}
\frac{\Delta {\cal T}}{\cal T} (\omega, T) = -\frac{3}{2\pi a^2}
\left[\sigma(\omega, T) - \sigma(\omega, T_{\rm amb})\right], 
\end{equation}
$T_{\rm amb}$ being the ambient temperature.
The relation \eqref{transmission} between transmission and absorption holds
provided that the reflectivity of the sample can be neglected, which is the case
in most of the experimental setups. 
The differential transmission can be viewed in two different ways. For a fixed time
delay between the pump and the probe pulses, it is sensitive to the energy
provided by the pump laser which is transferred to the electronic environment
via the surface plasmon, and thus to the temperature of the heat bath.
Alternatively, for a given pump intensity, increasing the time delay between the pump and the
probe scans the relaxation process of the electronic system as the bath
temperature decreases.

\begin{figure}[!t]
\begin{center}
\includegraphics[width=8truecm]{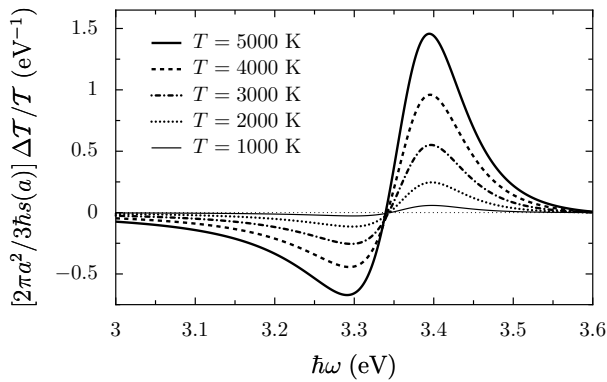}
\caption{\label{fig_transmission} Differential transmission
$\Delta {\cal T}/{\cal T}$  as a function of the probe energy
$\hbar \omega$ for increasing temperatures, resulting 
from increasing pump intensities and a fixed delay between pump and probe (or by
decreasing the time delay at a fixed pump intensity).
The presented results are for a sodium 
nanoparticle with a radius $a=\unit[2]{nm}$.}
\end{center}
\end{figure}

In Fig.~\ref{fig_transmission}, we present the differential transmission of
\eqref{transmission} for a sodium nanoparticle of radius $a=\unit[2]{nm}$,
which has a strong surface plasmon resonance around $\unit[3.5]{eV}$. We see that as the
temperature of the electronic system increases, $\Delta {\cal T}/{\cal T}$ becomes 
more and more pronounced since both, the linewidth and the redshift of the resonance 
frequency in \eqref{Breit-Wigner} increase with temperature. Similar results are observed 
in experiments, \cite{bigot, delfatti} where the amplitude of $\Delta {\cal T}/{\cal T}$  
is observed to decrease as a function of the time delay between the pump and the probe laser 
fields. This is accompanied by the blueshift of the crossing of the differential
transmission curves with the zero line as the time delay increases. Moreover,
the asymmetry of $\Delta{\cal T}/{\cal T}$ is observed in the experiments and
obtained in our calculations.

Our results could provide a possibility to fit the experimental results on 
metallic nanoparticles excited by a pump laser field in order to extract the
temperature, and thus could assist in analyzing the relaxation process.

%===========================================================================
%===========================================================================
%===========================================================================
%===========================================================================
\section{Conclusion}
\label{sec_ccl}
The role of the electronic environment on the surface plasmon resonance in 
metallic nanoparticles has been analyzed in this work. By means of a separation 
into collective and relative coordinates for the electronic system, we have
shown that the coupling to particle-hole excitations leads to a finite lifetime 
as well as a frequency shift 
of the collective surface plasmon excitation. The size and temperature dependence of the 
linewidth of the surface plasmon has been investigated by means of a semiclassical 
evaluation within the mean-field approximation, together with a low-temperature 
expansion. In addition to the well-known size dependence of the linewidth, 
we have demonstrated that an increase in temperature
leads to an increasing width of the resonance. The effect of finite temperature has
been found to be weak, in qualitative agreement with the experimental results.

We have analyzed the spill-out effect arising from the electron density outside
the nanoparticle. Our semiclassical analysis has led to a good
agreement with LDA calculations. In order to achieve this, it was necessary to
introduce an effective radius of the nanoparticle which accounts for the details
of the self-consistent mean-field potential. The ratio of spill-out electrons
over the total number as well as the resulting redshift of the surface plasmon
frequency scale inversely with the size of the nanoparticle.
The spill-out-induced redshift was shown, by means of a
Sommerfeld expansion, to increase with the temperature.

We have demonstrated that the coupling between the electronic center of mass and the 
relative coordinates results in an additional redshift of the surface 
plasmon frequency. This effect is of the same order as the redshift induced by 
the spill-out effect and presents a similar size and temperature dependence.
Thus it has to be taken into account in the description of numerical and experimental 
results. Our semiclassical theory predicts that for the smallest sizes of nanoparticles, 
the environment-induced redshift should exhibit a nonmonotonic behavior as a function of 
the size, as confirmed by numerical calculations. This is not the case for the redshift 
caused by the spill-out effect, and thus permits to distinguish between the two effects.

Our theory of the thermal broadening of the surface plasmon 
resonance, together with the temperature dependence of the resonance frequency, 
qualitatively explains the observed differential transmission that one measures 
in time-resolved pump-probe experiments. Our findings could open the possibility 
to analyze relaxation processes in excited nanoparticles.

%===========================================================================
%===========================================================================
%===========================================================================
%===========================================================================
\begin{acknowledgments}
We thank J.-Y.~Bigot, F.~Gautier, V.~Halt\'e, and
P.-A.~Hervieux for useful discussions. 
We acknowledge financial support by DAAD and \'Egide through the Procope
program, as well as by the BFHZ-CCUFB.
\end{acknowledgments}

\appendix
%===========================================================================
%===========================================================================
%===========================================================================
%===========================================================================
\section{Low-temperature expansion for integrals involving two Fermi functions}
In this appendix, we calculate the function $\cal F$ defined in
\eqref{F}, which involves the integral of two Fermi distributions.  
Integrating \eqref{F} by parts yields
\begin{equation}
\label{int_byparts}
{\cal F}(\mu, \hbar \omega) = \int_{\hbar \omega}^\infty {\rm
d}\varepsilon 
\left(-\frac{{\rm d} F}{{\rm d} \varepsilon}\right)
{\cal H}\left(\frac{\varepsilon}{\hbar \omega}\right)
\end{equation}
with
\begin{equation}
{\cal H}(x) = \int_1^x {\rm d} x' H(x'), 
\end{equation}
the function $H$ being defined in \eqref{H_function}.
In \eqref{int_byparts},
we have introduced the notation 
$F(\varepsilon)=\left[1-f(\varepsilon)\right]f(\varepsilon-\hbar \omega)$.
In the low-temperature limit $\hbar\omega\gg k_{\rm B}T$, we get
\begin{equation}
-\frac{{\rm d} F}{{\rm d} \varepsilon} \approx 
-\beta \frac{{\rm e}^{\beta(\varepsilon-\mu)}}{\left[{\rm
e}^{\beta(\varepsilon-\mu)}+1\right]^2} 
+\beta \frac{{\rm e}^{\beta(\varepsilon-\hbar \omega-\mu)}}{\left[{\rm
e}^{\beta(\varepsilon-\hbar \omega-\mu)}+1\right]^2}, 
\end{equation}
which corresponds to two peaks of opposite sign centered at $\varepsilon = \mu$ 
and at $\varepsilon = \mu+\hbar \omega$. It is therefore helpful to expand in
\eqref{int_byparts} the 
function ${\cal H}$ around $\varepsilon = \mu$ and $\varepsilon = \mu+\hbar\omega$. 
For low temperatures and for $|\hbar\omega-\mu|\gtrsim k_{\rm B}T$, we obtain
\begin{align}
\label{F_inter}
&{\cal F}(\mu, \hbar \omega)\simeq
{\cal H}\left(1+\frac{\mu}{\hbar\omega}\right)
-{\cal H}\left(\frac{\mu}{\hbar\omega}\right)\Theta(\mu-\hbar\omega)\\
&+\frac{\pi^2}{6}\left(\frac{k_{\rm B}T}{\hbar\omega}\right)^2
\left[H'\left(1+\frac{\mu}{\hbar\omega}\right)
-H'\left(\frac{\mu}{\hbar\omega}\right)
\Theta\left(\mu-\hbar\omega\right) \right] \nonumber
\end{align}
where $H'$ denotes the derivative of $H$. 
This treatment is appropriate for all values of $\omega$, except in a range of
order $k_{\rm B}T$ around $\mu$. There, the nonanalyticity
of $\cal H$ has to be properly accounted for\cite{thesis} since
for $x$ near $1$, ${\cal H}(x)\simeq(16/15)(x-1)^{5/2}$. 
At $\hbar\omega=\mu$, an additional term thus appears, and 
\begin{equation}
\label{T_52}
{\cal F}(\mu, \mu)\simeq
{\cal H}(2)
+\frac{\pi^2}{6}\left(\frac{k_{\rm B}T}{\hbar\omega}\right)^2 H'(2)
-C\left(\frac{k_{\rm B}T}{\hbar\omega}\right)^{5/2}
\end{equation}
with 
\begin{equation}
C=
\int_0^\infty{\rm d}x \frac{{\rm e}^x x^{5/2}}{\left({\rm e}^x+1\right)^2}
\approx3.07.
\end{equation}

In order to pursue the evaluation of \eqref{F_inter}, we need to determine the 
chemical potential $\mu$. Since the DOS of \eqref{DOS_0}, once summed
over $l$ and $m$ yields \cite{weick} the three-dimensional bulk DOS proportional
to $\sqrt{\varepsilon}$, we can use the standard Sommerfeld expression for 
the chemical potential of free fermions. \cite{ashcroft} We thus get
for $|\hbar\omega-\mu|\gtrsim k_{\rm B}T$
\begin{align}
\label{somm-like_result}
{\cal F}(\mu, \hbar \omega)&\simeq
\int_{\max{(\varepsilon_{\rm F}, \hbar\omega)}}^{\varepsilon_{\rm
F}+\hbar \omega}
\frac{{\rm d} \varepsilon}{\hbar \omega}
H\left(\frac{\varepsilon}{\hbar\omega}\right)
+ \frac{\pi^2}{6} \frac{\varepsilon_{\rm F}}{\hbar\omega} \left(\frac{T}{T_{\rm F}}\right)^2
\nonumber\\
&\times\bigg\{\frac{\varepsilon_{\rm F}}{\hbar\omega}\left[
H'\left(1+\frac{\varepsilon_{\rm F}}{\hbar \omega}\right)
-H'\left(\frac{\varepsilon_{\rm F}}{\hbar \omega}\right)
\Theta\left(\varepsilon_{\rm F}-\hbar\omega\right) \right]  \nonumber\\
&\hspace{.4truecm}-\frac 12 \left[H\left(1+\frac{\varepsilon_{\rm F}}{\hbar\omega}\right)-
H\left(\frac{\varepsilon_{\rm F}}{\hbar \omega}\right)
\Theta\left(\varepsilon_{\rm F}-\hbar \omega\right) 
\right] \bigg\},
\end{align}
where $T_{\rm F}$ is the Fermi temperature.
Inserting the result \eqref{somm-like_result} into \eqref{gamma_T_inter}, 
and using the expression \eqref{H_function} for $H$, we finally obtain
\eqref{Sigma}.

%===========================================================================
%===========================================================================
%===========================================================================
%===========================================================================


\begin{thebibliography}{}
\bibitem{footnote_T}
The energy required to ionize the nanoparticle 
is the work function plus the charging energy. This corresponds to a
temperature of several $\unit[10^5]{K}$. Therefore for reasonable temperatures,
nanoparticles cannot be thermoionized.
\bibitem{HF}
Within the jellium model, the LDA is not responsible for any underestimation of the number of
spill-out electrons, as it has been shown using Hartree-Fock and RPA
calculations. \cite{brack} Our results show that the frequency shift induced by the
electronic environment has to be taken into account in order to explain the
above-mentioned discrepancy.
\bibitem{footnote_TDLDA}
We can compare the semiclassical and numerical results since both approaches are
based on the many-body Hamiltonian \eqref{H}, and thus contain the same physical
ingredients.
\end{thebibliography}
\end{document}